\documentclass{ws-ijmpa}
\usepackage{amsmath}
\usepackage{graphicx}
\def\ast{\alpha^*}

\def\pdellx'{\frac{\partial}{\partial x'}}
\def\pdellw'{\frac{\partial}{\partial w'}}

\newcommand{\be}{\begin{equation}}
\newcommand{\ee}{\end{equation}}
\def\bed{\begin{displaymath}}
\def\eed{\end{displaymath}}
\def\bea{\begin{eqnarray}}
\def\eea{\end{eqncrray}}
\def\[{$$}
\def\]{$$}
\begin{document}

\title{The study of Nuclear binding energy for $A\geq100$ based on Odd-Even staggering of nuclear masses}

\author{B. B. Jiao	\footnote{E-mail:baobaojiao91@126.com}\\
	School of Nuclear Science and Engineering, East China University of Technology\\ Nanchang 330013, People's Republic of China\\}

\date{May 10, 2020}

\maketitle
{\small Accurate measurement of nuclear masses plays a key role in the nuclear physics, nuclear technology and astrophysical fields, especially in the calculation of nucleosynthesis and fast neutron capture processes.
The existing nuclear masses formula and nuclear masses model has undoubtedly achieved very good results, but it is still not satisfactory for some nuclear masses (especially near the neutron drip line), and even many nuclear masses have no prediction.
Although there are many studies in Odd-Even staggering (OES) of nuclear masses, but the research on nuclear masses by using the systematicness of OES is indeed very few.
Our purpose in this paper is to describe an empirical formula for Odd-Even staggering of nuclear masses that can be useful in describing and predicting nuclear masses.
We empirically obtained the formula of odd-$Z$ (odd-$N$) nuclei and even-$Z$ (even-$N$) nuclei based on studying the OES of nuclear masses (AME2012), where $Z$ and $N$ represent the number of proton and neutron.
Then describe and predict the nuclear masses with mass number $A\geq100$.
With the proton (neutron) empirical pairing gap from the OES of the binding energies and AME2012 database, the root-mean-square deviation of even-$Z$ nuclei and odd-$Z$ nuclei that we have successfully obtained 208 keV and 238 keV, respectively.
The RMSD of even-$N$ nuclei and odd-$N$ nuclei is 222 keV and 240 keV.
The result shows that our predicted values are compared well with values in AME2016, and some predicted values agree better with the experimental values.
These results demonstrate that our empirical formulas have good accuracy and reliability.
Another advantage of these formulas is that they use less known nuclear masses to predict unknown nuclear masses.
In addition, this paper also uses BP neural network to study proton Odd-Even staggering of nuclear masses (even-$Z$ and odd-$Z$ nuclei) and neutron Odd-Even staggering of nuclear masses (even-$N$ and odd-$N$ nuclei).
The RMSD of even-$Z$ and odd-$Z$ nuclei is 141 keV and 159 keV; the RMSD of even-$N$ and odd-$N$ nuclei is 150 keV and 160 keV.
The results show that the RMSD of nuclear masses based on neural network 60-80 keV decrease than that based on empirical formula (the accuracy is increased by about 32\%).
Accurate nuclear mass is helpful to the research of nuclear physics, nuclear technology and astrophysics.}

keywords: nuclear masses; Odd-Even staggering of nuclear masses; neural network.


\section{Introduction}

Nuclear masses has attracted much attention[1-22].
Precise prediction and measurement of nuclear mass have always been an important issue in nuclear physics and astrophysics.
It is found that the static mass of the nucleus is always less than the sum of the mass of the nucleon that makes up the nucleus,
the difference is the mass excess.
The study of nuclear mass based on global mass relation$^{[1-11]}$and regional mass relation$^{[14-22]}$is of great concern.
At present, the comprehensive databases are AME2003$^{[14]}$, AME2012$^{[15]}$ and AME2016$^{[16]}$.

The phenomenon of pair correlation that people notice is the Odd-Even staggering of nuclear masses.
In recent years, more and more people have paid close attention to the study of Odd-Even staggering of nuclear masses$^{[23-26]}$, but few people use Odd-Even staggering of nuclear masses to systematically describe and predict nuclear masses.
This paper describes and predicts the nuclear masses based on the systematic study of Odd-Even staggering of nuclear masses.
We obtained the empirical formula of odd-$Z$ (odd-$N$) nuclei and even-$Z$ (even-$N$) nuclei based on the Odd-Even staggering of nuclear masses by AME2012 database, and then obtained the nuclear masses of calculation.
The RMSD of even-$Z$ nuclei and odd-$Z$ nuclei is 208 keV and 238 keV;
the RMSD of even-$N$ nuclei and odd-$N$ nuclei is 222 keV and 240 keV, respectively.
In addition, there are many papers using artificial neural networks in nuclear physics [27-37] and other subjects [38-40].
In the 1990s, people have used neural networks [27] to predict the mass of atomic nuclei.
Research in recent years, many improvements have been made based on the neural network approach to reduce the deviation of the calculated values or the predicted values [34-37].
Ref. [34] shows that the accuracy of the Duflo-Zuker mass formula is improved by using the Bayesian neural network approach, the RMSD is reduced from 503 keV to 286 keV (the accuracy is increased by about 40\% );
Ref. [37] used Levenberg-Marquardt neural network approach to study the nuclear masses in AME2012 database, results show that Levenberg-Marquardt neural network method is helpful to improve the accuracy of mass models, for a simple liquid drop formula: the RMSD between the predicted value and the 2353 experimental known masses decreased sharply from 2.455 MeV to 0.235 MeV, while for some other mass models, the accuracy is improved by about 30\%.
This paper we use BP neural network  to study the Odd-Even staggering of nuclear masses (AME2012 database).
Results show that the RMSD based on the neural network combined with Odd-Even staggering of nuclear masses is 60-80 keV less than that based on empirical formula (the accuracy is increased by about 32\%).
The RMSD of even-$Z$ nuclei and odd-$Z$ nuclei is 141 keV and 159 keV;
the RMSD of even-$N$ nuclei and odd-$N$ nuclei is 150 keV and 160 keV, respectively.

In this paper, we use Odd-Even staggering of nuclear masses combined with AME2012 database to study the nuclear masses of $A\geq100$.
In Sec. 2, we use the known nuclear mass in ame2012 database and the three parameters Odd-Even staggering of nuclear masses formula to get many data sets of nuclear masses.
The empirical formulas are obtained based on the selected nuclei, and then calculated the nuclei with known mass.
In Sec. 3, we used BP neural network to study the Odd-Even staggering of nuclear masses, and then obtained the RMSD of known masses nuclei based on BP neural network and databases.
In Sec. 4, compares the predicted value calculated by empirical formulas (BP neural network method) and AME2012 database with the experimental value in AME2016, which shows that the predicted value in this paper is close to the experimental value.
In Sec. 5, discusses and summarizes this article.

\section{The method of empirical formula}
There are several kinds of pairing gap parameters$^{[23-26]}$, here we study the three-point formula $\Delta_p^{(3)}$ and $\Delta_n^{(3)}$ for proton pairing gaps and neutron pairing gaps,

\begin{eqnarray}
\label{eq:1}
\Delta_p^{(3)}(Z,N)=\frac{1}{2}[B(Z+1,N)-2B(Z,N)+B(Z-1,N)]\nonumber\\[1mm]
=\frac{1}{2}[S_p(Z+1,N)-S_p(Z,N)].\nonumber\\[1mm]
\end{eqnarray}

\begin{eqnarray}
\label{eq:2}
\Delta_n^{(3)}(Z,N)=\frac{1}{2}[B(Z,N+1)-2B(Z,N)+B(Z,N-1)]\nonumber\\[1mm]
\ =\frac{1}{2}[S_n(Z,N+1)-S_n(Z,N)].\nonumber\\[1mm]
\end{eqnarray}
where $B(Z,N)$ denotes the binding energy of the $(Z, N)$ nucleus with $A = Z + N$.

Here we define the binding energy as a positive value, which is easy to get:
\begin{eqnarray}
\label{eq:3}
\Delta_p^{(3)}(Z,N)=\frac{1}{2}[2M(Z,N)-M(Z-1,N)-M(Z+1,N)].\nonumber\\[1mm]
\end{eqnarray}

\begin{eqnarray}
\label{eq:4}
\Delta_n^{(3)}(Z,N)=\frac{1}{2}[2M(Z,N)-M(Z,N-1)-M(Z,N+1)].\nonumber\\[1mm]
\end{eqnarray}
The experimental nuclear mass is usually determined from the known atomic mass in AME databases. However, electron binding energy and Coulomb energy are usually neglected in nuclear mass studies.
It can be seen from eqs. (3) and (4) that the electron mass does not affect the calculation of OES of nuclear masses.
Therefore, we assume that the atomic mass is equal to the nuclear mass in this section.

As is depicted in Fig. 1, the OES of nuclear masses for even-$Z$ nuclei is less than zero and the OES of nuclear masses for odd-$Z$ nuclei is greater than zero.
It can be seen from Fig.1 that the black circles are different from red circles and green circles,
so the OES of nuclear masses for nuclei with $Z = 50, 82 $ are not included.
In addition, the OES of nuclear masses has certain linear characteristics.
Based on this behavior, we obtained the $\Delta_p^{(3)}$ formulas of even-$Z$ and odd-$Z$ nuclei for $A\geq100$:
\begin{eqnarray}
\label{eq:5}
\overline{\Delta_{p-even}^{(3)}}(A) \simeq \frac{-39600\cdot\mathrm{ln}A}{A}\ \mathrm{keV,} \nonumber\\[1mm]
\ \overline{\Delta_{p-odd}^{(3)}}(A) \simeq \frac{26000\cdot\mathrm{ln}A}{A}\ \mathrm{keV}.
\end{eqnarray}

\begin{figure}
\begin{center}
  \includegraphics[width=0.6\textwidth]{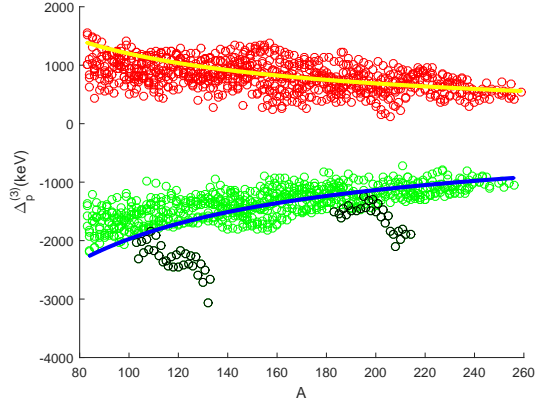}
\caption{(Color online) The odd-even staggering of nuclear masses for proton. We present separately for nuclei with even-$Z$ (green circles) and odd-$Z$ (red circles). The black circles represent the OES of nuclear masses with $Z = 50, 82 $ (the nuclei with a magic number of protons).
The yellow and blue curve are plotted in terms of Eq. (5).}
\label{fig:1}    
\end{center}   
\end{figure}

Fig. 2 shows that the OES of nuclear masses for even-$N$ nuclei is less than zero and the OES of nuclear masses for odd-$N$ nuclei is greater than zero.
In Fig.2, the black circles are different from red circles and green circles,
so the OES of nuclear masses for nuclei with $N=82, 126$ are not included.
In addition, the OES of nuclear masses has certain linear characteristics, then we obtained the $\Delta_n^{(3)}$ formulas of even-$N$ and odd-$N$ nuclei for $A\geq100$:

\begin{eqnarray}
\label{eq:6}
\overline{\Delta_{n-even}^{(3)}}(A) \simeq \frac{-32600\cdot\mathrm{ln}A}{A}\ \mathrm{keV,} \nonumber\\[1mm]
\ \overline{\Delta_{n-odd}^{(3)}}(A) \simeq \frac{26000\cdot\mathrm{ln}A}{A}\ \mathrm{keV}.
\end{eqnarray}

\begin{figure}
\begin{center}
  \includegraphics[width=0.6\textwidth]{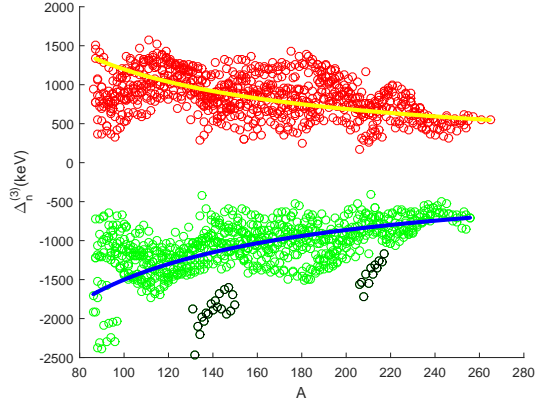}
\caption{(Color online) The odd-even staggering of nuclear masses for neutron. We present separately for nuclei with even-$Z$ (green circles) and odd-$Z$ (red circles), where black circles represent the OES of nuclear masses with $N=82, 126$ (the nuclei with a magic number of neutrons). The yellow and blue curve are plotted in terms of Eq. (6).}
\label{fig:2}      
\end{center} 
\end{figure}

Interestingly, the empirical formula of odd-$Z $ nuclei and odd-$N$ nuclei are on the same.
In addition, then we can easy get the formula of nuclear masses based on Eqs. (3) and (4).

\begin{eqnarray}
\label{eq:7}
M(Z,N)=\frac{2\Delta_p^{(3)}(A)+M(Z-1,N)+M(Z+1,N)}{2},\nonumber\\[1mm]
\end{eqnarray}

\begin{eqnarray}
\label{eq:8}
M(Z,N)=\frac{2\Delta_n^{(3)}(A)+M(Z,N-1)+M(Z,N+1)}{2}.\nonumber\\[1mm]
\end{eqnarray}
Result shows that the RMSD of even-$Z$ and odd-$Z$ nuclei is 208 keV and 238 keV;
the RMSD of even-$N$ and odd-$N$ nuclei is 222 keV and 240 keV, respectively.
Calculated values confirm that our new formula can be used to calculate and predict nuclear masses.

The nuclear mass is equal to the atomic mass minus masses of electrons and Coulomb energy plus the electron binding energy.
However, the electron binding energy, the electron mass and the Coulomb energy are often neglected in the research of nuclear masses.
But in this section, the role of the electron binding energy and the Coulomb energy are studied.
Because the electron mass has no effect on the odd even difference, therefore the electron mass is neglected in our studies.
The formula of nuclear masses is given by[46]: $M^{*}(Z, N)=M(Z, N)+B_{e}(Z)-B_{e}^{fit}(Z, A)$.
Where the formulas of the electron binding energy ($B_{e}$) and the Coulomb energy ($B_{e}^{fit}$) derived from [15] are respectively

\begin{eqnarray}
\label{eq:9}
B_{e}(Z)=14.4381\cdot Z^{2.39}+1.55468\cdot 10^{-6}\cdot Z^{5.35}\ \mathrm{eV},
\end{eqnarray}

\begin{eqnarray}
\label{eq:10}
B_{e}^{fit}(Z, A)= 505\cdot \frac{Z^2}{A^{1/3}}\cdot(1-0.76\cdot Z^{-2/3})\ \mathrm{eV}.
\end{eqnarray}

It can be seen from Eqs. (7) and (8) that the electron binding energy and Coulomb energy are only significant for the proton OES of nuclear masses.
In this section, we introduce the electron binding energy and Coulomb energy to obtain the new nuclear masses, and then obtain the known nuclear masses based on formula (7).
Fig. 3 and Fig. 4 represents the RMSD of the masses of even-$Z$ and odd-$Z$ nuclei, respectively .
We assume that the atomic mass is equal to the nuclear mass, by using the empirical formula combination (7) and even-$Z$ (odd-$Z$) OES obtained the RMSD of the known nuclear mass is $\sigma_1$ ($\sigma_5$);
when the binding energy of electron is considered, the RMSD is $\sigma_2$ ($\sigma_6$);
the RMSD obtained by considering Coulomb energy in nuclear mass is $\sigma_3$ ($\sigma_7$);
when the nuclear mass is obtained after used the electron binding energy to minus the Coulomb energy, and then obtained the RMSD is $\sigma_4$ ($\sigma_8$).

\begin{figure}
\begin{center}
  \includegraphics[width=0.6\textwidth]{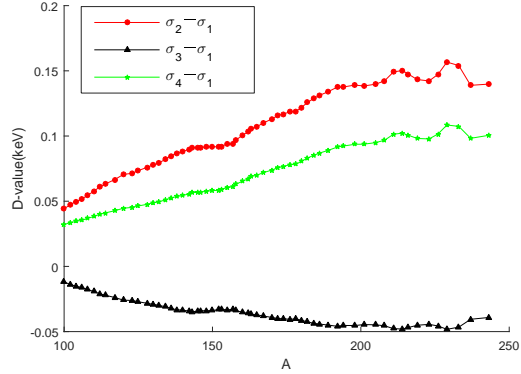}
\end{center}
\caption{(Color online) The difference of RMSD for even-$Z$ nuclei.}
\label{fig:3}       
\end{figure}

\begin{figure}
\begin{center}
  \includegraphics[width=0.6\textwidth]{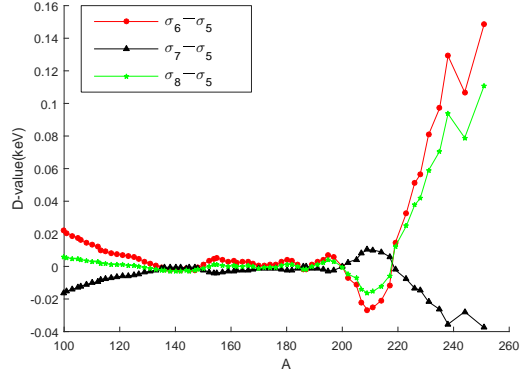}
\end{center}
\caption{(Color online) The difference of RMSD for odd-$Z$ nuclei.}
\label{fig:4}       
\end{figure}

The points in Fig. 3 represent the RMSD of nuclear masses, and the number of nuclei between the two points is 10.
Red dot line represents the difference of RMSD  between the $\sigma_2$ and $\sigma_1$.
The difference is greater than 0 means that the introduction of electron binding energy leads to the increase of the deviation, and the larger nuclear mass, the greater deviation will be obtained.
The black dot line represents the difference of RMSD  between the $\sigma_3$ and $\sigma_1$. The difference is less than 0 means that the increase of Coulomb energy reduces the deviation, and the larger nuclear mass, the smaller deviation will be obtained.
The green solid rectangular dotted line  represents the difference of RMSD  between the $\sigma_4$ and $\sigma_1$. The difference is greater than 0 and less than the red dotted line indicates means that the Coulomb energy offsets part of the electron binding energy.
However, the introduction of these two energies will cause the deviation to increase.

\section{The method of BP neural network}
In addition, our paper using the 'newff' function provided by the neural network toolbox of MATLAB 2015b to create the forward BP neural network, and then to study the OES of nuclear masses.
We study the OES of nuclear masses based on the neural network function of tansig [$f(x)=2/(1+e^{-2x})-1$],
which the function comes from the neural network toolbox in MATLAB 2015b.
The data sets Z, N, $ \Delta_p^{(3)}(Z,N)$ [or $\Delta_n^{(3)}(Z,N)]$ (calculated values of the OES of nuclear masses for known mass) as the input sample(that is training data) for the network.
After training obtained the residual interaction $ \Delta_p^{(3)}(Z,N)$ [or $\Delta_n^{(3)}(Z,N)]$ [analog values of the OES of nuclear masses for known mass] and $ \Delta_p^{(3)}(Z1,N1)$ [or $\Delta_n^{(3)}(Z1,N1)]$ [predicted value of the OES of nuclear masses for unknown mass].
Using OES of nuclear masses and Eqs. (7) and (8), combined with AME2012 and AME2016, the values of the known nuclear masses were calculated and the values of unknown nuclear masses were predicted. Then, the RMSD between the calculated values of nuclear mass and the corresponding experimental values in AME2012 and AME2016 was obtained.

In this section, we study the OES of even-$Z$ and odd-$Z$ nuclear masses respectively.
Using BP neural network got the fitting values of OES of nuclear masses, then obtained the calculated values of known masses.
Fig. 5 shows the RMSD between the calculated values of even-$Z$ (odd-$Z$) nuclear masses and the experimental values in AME2012 (AME2016) database.
The line in black (blue) represents the RMSD of comparing the calculated values (we obtained the calculated values based on AME2012) of even-$Z$ (odd-$Z$) nuclei masses with the experimental values in AME2012 database.
The curve in red(pink) is plotted by using the RMSD between calculated values (we obtained the calculated values based on AME2016) of even-$Z$ (odd-$Z$) nuclei and experimental values in AME2016 database.
The results show that the accuracy of the OES of nuclear masses is significantly improved by using BP neural network approach, the RMSD relative to experiment is reduced about 70 keV.

\begin{figure}
\begin{center}
  \includegraphics[width=0.6\textwidth]{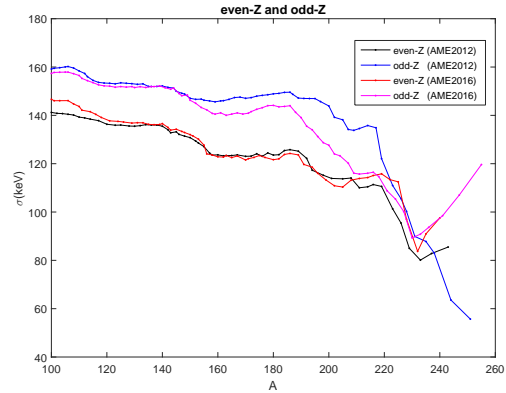}
\end{center}
\caption{(Color online) The RMSD for even-$Z$ and odd-$Z$ nuclei.}
\label{fig:5}       
\end{figure}

In Fig. 6, the lines represent the RMSD between the calculated values of even-$N$ (odd-$N$) nuclear masses and the experimental values in AME2012 (AME2016) database.
We use the black (blue) line represents the RMSD between calculated values (we obtained the calculated values based on AME2012) of even-$N$(odd-$N$) nuclear masses and experimental values in AME2012 database.
The curve in red(pink) corresponds to the RMSD between calculated values (we obtained the calculated values based on AME2016) of even-$N$ (odd-$N$) nuclei and experimental values in AME2016 database.
The result shows that the accuracy of the empirical formula of OES is significantly improved by using BP neural network approach, the RMSD relative to experiment is reduced about 70 keV.

\begin{figure}
\begin{center}
  \includegraphics[width=0.6\textwidth]{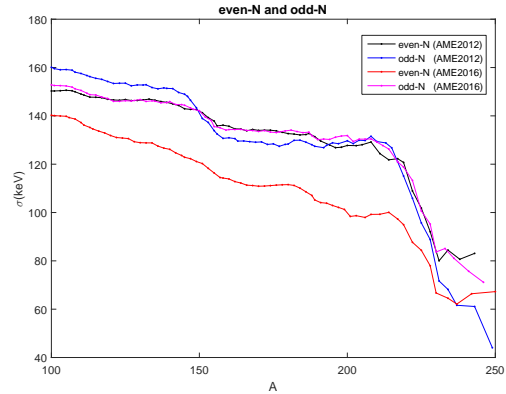}
\end{center}
\caption{(Color online) The RMSD for even-$N$ and odd-$N$ nuclei.}
\label{fig:6}       
\end{figure}

Figs. 5 and 6 show that the calculated values of nuclear masses based on BP neural network is better than that using empirical formulas, the accuracy is increased by about 32\%.
The accuracy of nuclear masses based on the proposed BP neural network method is better than that using empirical formulas, which we can also use Eqs. (7) and (8)  combination BP neural network obtained the predicted values of nuclear masses.

\section{Prediction of nuclear masses}

In this section, we used the predictable formula of nuclear masses to predict the nuclear masses that are difficult to be measured experimentally.
The main idea of nuclear mass prediction is to get the $ \Delta_p^{(3)}(Z,N)$ and $\Delta_n^{(3)}(Z,N)$ of unknown mass, then calculate the unknown nuclear mass combined with Eq.(7) , Eq.(8) and AME2016 database.
The OES of nuclear masses of unknown mass nuclei is obtained based on the empirical formulas and BP neural network method.
In table 1 and 2 we present a set of values among our predicted values, then compare with experimental and predicted values in AME2016.

We obtained the predicted values of unknown mass based on $ \Delta_p^{(3)}(Z,N)$ [Eq. (5)] and Eq.(7).
In Table 1, we give 38 predicted values in comparison with the dates in AME2016.
ME$_{1}$ and ME$_{2}$ are obtained based on empirical formula and BP neural network method respectively.
The dev0 represents the deviation of mass excess in AME2016 database;
dev1 is the difference between the values (ME2016) of the mass excess in the AME2016 database and our predicted values (ME$_{1}$);
we use dev2 to represents the difference between the values (ME2016) and our predicted values (ME$_{2}$).
There are five predicted values($^{141}I$, $^{190}Tl$, $^{194}Bi$, $^{198}At$, $^{202}Fr$) have their experimental values in AME2016 database.
We found that our predicted values of simulation is according with experiment.
The predicted values we get also agree with those in the AME2016 database, especially for those heavy nuclei and superheavy nuclei.
In addition, the RMSD between our calculated values (ME$_{1}$) and the AME2016 database (ME2016) is 182keV, and the RMSD between ME2016 and ME$_{2}$ is 167keV.
It can be seen that the deviation of predicted values based on neural network is smaller than that based on empirical formulas.

%
\begin{table}
\centering
\caption{The difference between nuclear masses in AME2016 database and the predicted values [obtained by the OES of even-$Z$ (odd-$Z$) nuclei and AME2012]. (keV)}
\label{tab:1}       
\begin{tabular}{lllllll}
\hline\noalign{\smallskip}
Nucl.     & ME2016    &  dev0 &  ME$_{1}$ &  ME$_{2}$    & dev1   & dev2  \\
\noalign{\smallskip}\hline\noalign{\smallskip}
$^{101}In$	&	-68610	&	200	&	-68372	&	-68700	&	-238	&	90	\\	
$^{114}I$	&	-72800	&	150	&	-72418	&	-72455	&	-382	&	-345	\\	
$^{116}Cs$	&	-62040	&	100	&	-62070	&	-62031	&	30	&	-9	\\	
$^{139}Gd$	&	-57630	&	200	&	-57521	&	-57745	&	-109	&	115	\\	
$^{141}I$	&	\textbf{-59927}	&	16	    &	-59878	&	-60239	&	-49	&	312	\\	
$^{150}Tm$	&	-46490	&	200	&	-46771	&	-46341	&	281	&	-149	\\	
$^{167}Re$	&	-34830	&	40	&	-35138	&	-34976	&	308	&	146	\\	
$^{170}Ir$	&	-23360	&	90	&	-23308	&	-23073	&	-52	&	-287	\\	
$^{174}Au$	&	-14240	&	90	&	-14185	&	-13943	&	-55	&	-297	\\	
$^{178}Tl$	&	-4790	&	90	&	-4607	&	-4355	&	-183	&	-435	\\	
$^{178}Ta$	&	-50600	&	50	&	-50331	&	-50528	&	-269	&	-72	\\	
$^{185}Bi$	&	-2234	&	80	&	-2744	&	-2517	&	510	&	283	\\	
$^{190}Tl$	&	\textbf{-24372}&	8	&	-24211	&	-24137	&	-161	&	-235	\\	
$^{194}Bi$	&	\textbf{-16029}&	6	&	-15920	&	-15792	&	-109	&	-237	\\	
$^{198}At$	&	\textbf{-6715}&	6	&	-6732	&	-6517	&	17	&	-198	\\	
$^{202}Fr$	&	\textbf{3096}&	7	&	2983	&	3186	&	113	&	-90	\\	
$^{222}Pa$	&	22160	&	70	&	22023	&	22110	&	137	&	50	\\	
$^{232}Am$	&	43340	&	300	&	43401	&	43447	&	-61	&	-107	\\	
$^{232}Np$	&	37360	&	100	&	37542	&	37442	&	-182	&	-82	\\	
$^{233}Am$	&	43260	&	100	&	43154	&	43161	&	106	&	99	\\	
$^{236}Am$	&	46040	&	110	&	46319	&	46226	&	-279	&	-186	\\	
$^{237}Am$	&	46570	&	60	&	46776	&	46657	&	-206	&	-87	\\	
$^{239}Bk$	&	54250	&	210	&	54316	&	54277	&	-66	&	-27	\\	
$^{241}Bk$	&	56030	&	200	&	56150	&	56049	&	-120	&	-19	\\	
$^{245}Es$	&	66370	&	200	&	66420	&	66346	&	-50	&	24	\\	
$^{246}Am$	&	64994	&	18	&	64941	&	64950	&	53	&	44	\\	
$^{247}Am$	&	67150	&	100	&	66976	&	66980	&	174	&	170	\\	
$^{248}Es$	&	70300	&	50	&	70392	&	70265	&	-92	&	35	\\	
$^{248}Bk$	&	68080	&	70	&	68210	&	68176	&	-130	&	-96	\\	
$^{249}Es$	&	71180	&	30	&	71235	&	71112	&	-55	&	68	\\	
$^{250}Es$	&	73230	&	100	&	73416	&	73311	&	-186	&	-81	\\	
$^{252}Md$	&	80510	&	130	&	80729	&	80621	&	-219	&	-111	\\	
$^{252}Bk$	&	78540	&	200	&	78548	&	78548	&	-8	&	-8	\\	
$^{253}Md$	&	81170	&	30	&	81342	&	81230	&	-172	&	-60	\\	
$^{254}Md$	&	83450	&	100	&	83647	&	83545	&	-197	&	-95	\\	
$^{256}Md$	&	87460	&	120	&	87591	&	87544	&	-131	&	-84	\\	
$^{257}Lr$	&	92670	&	40	&	92646	&	92551	&	24	&	119	\\	
$^{263}Bh$	&	114500	&	310	&	114518	&	114519	&	-18	&	-19	\\	
\noalign{\smallskip}\hline
\end{tabular}
\end{table}

By using $\Delta_n^{(3)}(Z,N)$ [Eq. (6)] and Eq.(8) obtained the predicted values.
Table 2 lists 38 predicted values in comparison with the dates in the AME2016 database.
Where, ME$_{3}$ (ME$_{4}$) are obtained based on empirical formula (BP neural network method).
The dev0 is the deviation of mass excess in AME2016 database;
dev3 is the difference between the values (ME2016) of the mass excess in the AME2016 database and our predicted values (ME$_{3}$);
we use dev4 to represents the difference between the values (ME2016) and our predicted values (ME$_{4}$).
Table 2 shows that our predicted values are close to the data in AME2016 database,
and the prediction value is more accurate in the heavy nuclear region.
Moreover, the RMSD between ME2016 and ME$_{1}$ is 232keV, and the RMSD between ME2016 and ME$_{2}$ is 190keV.
Again, it is shown that the predicted value obtained by neural network is more accurate than that obtained by empirical formula.

\begin{table}
\centering
\caption{The difference between nuclear masses in AME2016 database and the predicted values [obtained by the OES of even-$N$ (odd-$ N$) nuclei and AME2012].(keV)}
\label{tab:2}       
\begin{tabular}{lllllll}
\hline\noalign{\smallskip}
Nucl.     & ME2016    &  dev0 &  ME$_{3}$ &  ME$_{4}$    & dev3   & dev4  \\
\noalign{\smallskip}\hline\noalign{\smallskip}
$^{101}In$	&	-68610	&	200	&	-68993	&	-68810	&	383	&	200	\\	
$^{114}I$	&	-72800	&	150	&	-72645	&	-72521	&	-155	&	-279	\\	
$^{118}Ba$	&	-62350	&	200	&	-62424	&	-62595	&	74	&	245	\\	
$^{129}Cd$	&	\textbf{-63058}&	17	&	-63405	&	-63734	&	347	&	676	\\	
$^{141}I$	&	\textbf{-59927}&	16	&	-60327	&	-59926	&	400	&	-1	\\	
$^{153}Yb$	&	-47210	&	200	&	-47269	&	-47261	&	59	&	51	\\	
$^{154}Lu$	&	-39720	&	200	&	-39634	&	-39595	&	-86	&	-125	\\	
$^{157}Hf$	&	-38900	&	200	&	-39145	&	-39131	&	245	&	231	\\	
$^{158}Ta$	&	-31170	&	200	&	-31208	&	-31193	&	38	&	23	\\	
$^{161}W$	&	-30560	&	200	&	-30865	&	-30801	&	305	&	241	\\	
$^{162}Re$	&	-22500	&	200	&	-22630	&	-22587	&	130	&	87	\\	
$^{165}Os$	&	-21800	&	200	&	-22148	&	-22016	&	348	&	216	\\	
$^{165}Tb$	&	-60570	&	100	&	-60989	&	-60677	&	419	&	107	\\	
$^{167}Re$	&	-34840	&	40	&	-34843	&	-35099	&	3	&	259	\\	
$^{169}Pt$	&	-12510	&	200	&	-12890	&	-12691	&	380	&	181	\\	
$^{170}Ir$	&	-23360	&	90	&	-23459	&	-23127	&	99	&	-233	\\	
$^{173}Hg$	&	-2710	&	200	&	-3101	&	-2840	&	391	&	130	\\	
$^{174}Au$	&	-14240	&	90	&	-14343	&	-13967	&	103	&	-273	\\	
$^{178}Tl$	&	-4790	&	90	&	-5043	&	-4621	&	253	&	-169	\\	
$^{178}Ta$	&	-50600	&	50	&	-50279	&	-50414	&	-321	&	-186	\\	
$^{182}Lu$	&	-41880	&	200	&	-41511	&	-41702	&	-369	&	-178	\\	
$^{185}Bi$	&	\textbf{-2240}&	80	&	-1889	&	-2100	&	-351	&	-140	\\	
$^{190}Tl$	&	\textbf{-24372}&	8	&	-24722	&	-24422	&	350	&	50	\\	
$^{194}Bi$	&	\textbf{-16029}&	6	&	-16241	&	-15877	&	212	&	-152	\\	
$^{198}At$	&	\textbf{-6715}&	6	&	-6887	&	-6505	&	172	&	-210	\\	
$^{198}Ir$	&	-25820	&	200	&	-25636	&	-25683	&	-184	&	-137	\\	
$^{202}Fr$	&	\textbf{3096}	&	7	&	2924	&	3276	&	172	&	-180	\\
$^{212}Tl$	&	-1550	&	200	&	-1488	&	-1497	&	-62	&	-53	\\	
$^{220}Pa$	&	20220	&	50	&	20098	&	20198	&	122	&	22	\\	
$^{222}Pa$	&	22160	&	70	&	21983	&	22096	&	177	&	64	\\	
$^{226}Np$	&	32780	&	90	&	32701	&	32834	&	79	&	-54	\\	
$^{232}Np$	&	37360	&	100	&	37400	&	37352	&	-40	&	8	\\	
$^{235}Cm$	&	48030	&	200	&	47895	&	47881	&	135	&	149	\\	
$^{241}Cf$	&	59330	&	170	&	59282	&	59227	&	48	&	103	\\	
$^{243}Cf$	&	60990	&	110	&	61023	&	60951	&	-33	&	39	\\	
$^{247}Fm$	&	71670	&	120	&	71625	&	71579	&	45	&	91	\\	
$^{248}Bk$	&	68080	&	70	&	68251	&	68162	&	-171	&	-82	\\	
$^{256}Md$	&	87460	&	120	&	87486	&	87484	&	-26	&	-24	\\	
\noalign{\smallskip}\hline
\end{tabular}
\end{table}

From table 1 and table 2 we can see that our predicted values are close to the experimental values and predicted values, and some nuclear mass deviations are only tens of keV.
Although the deviation of some nuclear masses larger than is desired, it has little effect on the overall prediction of nuclear masses.
Therefore, both empirical formula and neural network method can be used to predict nuclear masses.
In addition, the results show that the predicted values of unknown masses based on BP neural network are better than that using empirical formulas.
More accurate predictions could be readily made if the OES of nuclear masses was more accurate.

\section{Discussion and Conclusions}

In this paper we study the OES of nuclear masses for even-$Z$ and odd-$Z$ nuclei (even-N and odd-N nuclei), then obtained the calculated and predicted values of nuclear masses.
In Sec. 2, the empirical formulas obtained from studying the OES of nuclear masses.
We obtained the nuclear masses for $A\geq100$ by using the empirical formulas and AME databases.
Although the nuclear mass with large error exists, it does not affect the overall description and prediction effect.
In addition, in Sec. 3 we use BP neural network to study the OES of nuclear masses. The results show that the BP neural network is useful for described and predicted the nuclear masses.

Using the empirical formula of OES obtained the calculated and predicted values of nuclear masses.
The known nuclear mass with mass number $A\geq100$ is in good agreement with the experimental value.
The RMSD of even-$Z$ nuclei and odd-$Z$ nuclei is 208 keV and 238 keV, and the RMSD of even-$N$ nuclei and odd-$N$ nuclei is 222 keV and 240 keV.
At the same time, the research shows that the OES of even-$Z$ nuclei (even-$N$ nuclei) is better than that of odd-$Z$ nuclei (odd-$N$ nuclei), so the RMSD of even-$Z$ nuclei (even-$N$ nuclei) is less than that of odd-$Z$ nuclei (odd-$N$ nuclei).
It can be seen from table 1 and table 2 that the predicted value based on AME2012 is consistent with the value in AME2016 database, and the larger the mass number, the smaller the deviation.
In addition, it is feasible to describe and predict the OES of nuclear masses based on BP neural network.
The calculated value based on BP neural network is in good agreement with the experimental value, the RMSD of even-$Z$ and odd-$Z$ nuclei is 141 keV and 159 keV; the RMSD of even-$N$ and odd-$N$ nuclei is 150 keV and 160 keV.
At the same time, the prediction value based on the OES of nuclear masses and AME2012 database is close to the values in the AME2016 database.
Because the OES of even-$Z$ nuclei and even-$N$ nuclei is statistically good,
so the RMSD of the even-$Z$ nuclei (even-$N$ nuclei) is smaller than the odd-$Z$ nuclei (odd-$N$ nuclei).
It is found that the deviation of nuclear mass by BP neural network is 60-80keV less than that by empirical formulas, and the deviation is reduced by 32\%.
The advantage of BP neural networks method is to reduce the calculation and prediction deviation of nuclear masses, but the disadvantage is that we can not participate in the operation of neural networks.

The result shows that the nuclear mass can be described and predicted by using the formula of OES.
In addition, the nuclear mass calculated and predicted by using BP neural network to study the OES of nuclear masses is also in good agreement with the values in databases.
The number of nuclei involved in the description and prediction of nuclear mass is 2, less nuclear involvement makes extrapolation easier.
The more accurate the calculated and predicted values of the OES of nuclear masses are, the more accurate the nuclear mass will be.


\begin{thebibliography}{0}
\bibitem{lab1}
C.F. Von Weizs$\ddot{\mathrm{a}}$cker, Z. Phys {\bf 96}, 431 (1935).
\bibitem{lab2}
P. M$\ddot{\mathrm{o}}$ller et al., At. Data Nucl. Data Tables  {\bf 59}, 185 (1995).
\bibitem{lab3}
J. Duflo and A.P. Zuker, Phys. Rev. C {\bf 52}, R23 (1995).
\bibitem{lab4}
G.T. Garvey and I. Kelson, Phys. Rev. Lett. {\bf 16}, 197 (1966).
\bibitem{lab5}
L. Geng, H. Toki and J. Meng, Prog. Theor. Phys. {\bf 113}, 785 (2005).
\bibitem{lab6}
S. Goriely, N. Chamel and J.M. Pearson, Phys. Rev. Lett. {\bf 102}, 152503 (2009).
\bibitem{lab7}
P. M$\ddot{\mathrm{o}}$ller et al., Phys. Rev. Lett. {\bf 108}, 052501 (2012).
\bibitem{lab8}
C. Qi, J. Phys. G: Nucl. Par. {\bf 42}, 045104 (2015).
\bibitem{lab9}
N. Wang, Z. Liang, M. Liu et al., Phys. Rev. C {\bf 82}, 044304 (2010).
\bibitem{lab10}
S. Goriely, F. Tondeur and J.M. Pearson, At. Data Nucl. Data Tables {\bf 77}, 311 (2001).
\bibitem{lab11}
M. Bao, Z. He, Y. Lu et al., Phys. Rev. C {\bf 88}, 064325 (2013).
\bibitem{lab12}
B. Krusche, Eur. Phys. J. A {\bf26}, 7 (2005).
\bibitem{lab13}
J.M. Pearson, S. Goriely, M. Samyn, Eur. Phys. J. A {\bf15}, 13 (2002).
\bibitem{lab14}
G.J. Fu, H. Jiang, Y.M. Zhao et al., Phys. Rev. C {\bf 82}, 034304 (2012).
\bibitem{lab15}
H. Jiang, G.J. Fu, Y.M. Zhao et al., Phys. Rev. C {\bf 82}, 054317 (2010).
\bibitem{lab16}
D. Lunney, J.M. Pearson and C. Thibault, Rev. Mod. Phys. {\bf 75}, 1021 (2003).
\bibitem{lab17}
B.B. Jiao, Mod. Phys. Lett. A {\bf 32}, 1850156 (2018).
\bibitem{lab18}
Z. Wu, S.A. Changizi and C. Qi, Phys. Rev. C {\bf 93}, 034334 (2016).
\bibitem{lab19}
G. Audi, A.H. Wapstra and C. Thibault, Nucl. Phys. A {\bf 729}, 337 (2003).
\bibitem{lab20}
G. Audi, F.G. Kondev, M. Wang et al., Chin. Phys. C {\bf 41}, 030001 (2017).
\bibitem{lab21}
M. Wang, G. Audi, A.H. Wapstra et al., Chin. Phys. C {\bf 36}, 1603 (2012).
\bibitem{22}
B.B. Jiao, Sci Sin-Phys Mech Astron {\bf 48}, 052001 (in Chinese) (2018).
\bibitem{23}
J.W. Clark, H. Li, Int. J. Mod. Phys. B {\bf20}, 5015 (2006).
\bibitem{24}
N.J. Costiris, E. Mavrommatis, K.A. Gernoth et al., Phys. Rev. C {\bf80}, 044332 (2009).
\bibitem{25}
S. Akkoyun, T. Bayram, T. Turker, Radiat. Phys. Chem. {\bf96}, 186 (2014).
\bibitem{26}
S. Akkoyun, T. Bayram, Int. J. Mod. Phys. E {\bf23}, 1450064 (2014).
\bibitem{27}
S. Gazula, J.W. Clark and H. Bohr, Nucl. Phys. A  {\bf540}, 1 (1992).
\bibitem{28}
K.A. Gernoth K, J.W. Clark, J.S. Prater et al., Phys. Lett. B  {\bf300}, 1 (1993).
\bibitem{29}
T. Bayram, S. Akkoyun and S.O. Kara, Ann. Nucl. Energy {\bf63}, 172 (2014).
\bibitem{30}
S. Athanassopoulos, E. Mavrommatis, K.A. Gernoth et al., Nucl. Phys. A  {\bf743}, 222 (2004).
\bibitem{31}
L. Alvarez-Ruso, K.M. Graczyk and E. Saul-Sala, Phys. Rev. C {\bf 99}, 025204 (2019).
\bibitem{32}
N.J. Costiris, E. Mavrommatis, K.A. Gernoth et al., Phys. Rev. C {\bf80}, 044332 (2009).
\bibitem{33}
S. Akkoyun, T. Bayram and T. Turker, Radiat. Phys. Chem. {\bf96}, 186 (2014).
\bibitem{34}
R. Utama and J. Piekarewicz, Phys. Rev. C {\bf 96}, 044308 (2017).
\bibitem{35}
R. Utama and J. Piekarewicz, Phys. Rev. C {\bf 97}, 014306 (2018).
\bibitem{36}
Z.M. Niu and H.Z. Liang, Physics Letters B {\bf 778}, 48 (2018).
\bibitem{37}
H.F. Zhang, L.H. Wang, J.P. Yin et al., J. Phys. G: Nucl. Par. {\bf44}, 045110 (2017).
\bibitem{38}
J. He, X. Tang, P. Gong et al., Ann. Nucl. Energy {\bf 112}, 1 (2018).
\bibitem{39}
D. Ma, T. Zhou, J. Chen et al., Nucl. Eng. Des. {\bf 320}, 400 (2017).
\bibitem{40}
K.X. Peng, J.B. Yang, X.G. Tuo et al., Mod. Phys. Lett. B {\bf 30}, 87 (2016).

\end{thebibliography}
\end{document}